\documentstyle[epsfig,psfig]{texas}

\begin{document}

\title{High Energy Cosmic Neutrinos Astronomy: The ANTARES Project\footnote{Talk given at the 19$^{\text {th}}$ Texas Symposium, Paris, December 1998.}}
\author{Stephane BASA$^{1}$, on behalf of the ANTARES collaboration}
\address{(1) Centre de Physique des Particules de Marseille\\
163 Avenue de Luminy - Case 907\\
13288 Marseille Cedex 9, France\\
{\rm Email: basa@cppm.in2p3.fr}}

\begin{abstract}
Neutrinos may offer a unique opportunity to explore the far Universe at high energy. The ANTARES collaboration aims at building a large undersea neutrino detector able to observe astrophysical sources (AGNs, X-ray binary systems, ...) and to study particle physics topics (neutrino oscillation, ...).
After a description of the research opportunities of such a detector, a status report of the experiment will be made.
\end{abstract}

\section{Introduction}

Many questions are still open in Astronomy and in Particle Physics: what is the origin of the most energetic Cosmic Rays and of the Gamma Ray Bursts, what is the nature of the Dark Matter, do the neutrinos really oscillate as indicated by the SuperKamiokande result \cite{ref1}, ...

To give an answer to some of these questions, it is important to observe the Universe in a wide energy range and at long distance. To reach this goal an abundantly produced messenger is needed, it must able to bring informations out of the source, through long distances, until the detector. Among all the possibilities, only four particles have attracted the scientist's attention: the photon, the proton, the neutron and the neutrino.

  {\bf The photon} is the most natural and the favourite probe of the astronomers. Its observation over more than 18 orders of magnitude in energy has revolutionised our conception of the World. Unfortunately, at high energy ($>$ TeV), it interacts  with the less energetic ambient photons (infra-red, Cosmic Microwave Background and radio-wave photons) according to the pair-production mechanism:
  $$ \gamma \gamma \rightarrow e^+ e^-$$
  This so-called Greisen-Zatsepin-Kuzmin (GZK) cutoff prevents the observation of TeV photons above about 50 Mpc. This mechanism explains probably why only the nearest blazars have been observed in this energy range (the only extra-galactic sources observed so-far, Mkn 421 and Mkn 501, are also the two nearest blazars with z$\approx$0.03). Therefore the deep observation of the Universe at high energy is difficult with the photon. 
  
  {\bf The proton} is abundantly produced in the Universe and is therefore a good candidate. Unfortunately, at energies lower than a few EeV, it is deflected by the galactic and the extra-galactic magnetic field and its directional information is lost. At ultra high energy the tracing back of the sources is possible. However fluxes are very low and the proton becomes also sensitive to the GZK cutoff by interacting with the photons of the Cosmic Microwave Background (the mean free path is about z=1 for a 10 EeV proton and z=0.03 for a 100 EeV proton).
  
  {\bf The neutron} is electrically neutral and points back to the source. However its life time is too short. Even at ultra high energy ($>$ EeV), it can not travel more than 10 kpc, which is roughly the radius of our Galaxy.

  The last candidate is {\bf the neutrino}. It is probably the most promising: it is stable and neutral, allowing it to point back to its production point. Moreover, unlike the photon, it interacts only weakly with matter and photon fields. This last property leads to two major advantages and one drawback:
\begin{itemize}
  \item It can easily escape from very dense sources which trap all the others particles.
  \item It can travel over very large distances up to the detector.
  \item A very large detector will be necessary to observe it on Earth.
\end{itemize}
  
  Up to now this drawback has considerably limited its observation. The supernova SN1987A has been the first, and still the only, source of extraterrestrial neutrinos other than the sun. 

  The purpose of the ANTARES collaboration is to build a deep underwater neutrino detector. This one should be able to observe astrophysical sources and to study particle physics topics. After having discussed the scientific motivations and the constraints on the detector design, a status report on the project will be presented.

\section{Scientific Motivations}

\subsection{Sources of neutrinos}

The mechanisms leading to the production of neutrinos in the Universe can be classified into two categories: accelerating and non-accelerating mechanisms.

\subsubsection{Accelerating mechanisms}

Neutrinos can be produced according to the ``Beam Dump'' mechanism where an accelerated proton interacts with a target composed of protons and/or photons:
\begin{eqnarray}
\label{eq:nuprod}
p + \mathrm{target} \, (p,\gamma) \longrightarrow & \pi^{0} \quad , & \pi^{\pm} \quad  , \quad (\ldots)\\
 &\downarrow \quad \: & \: \downarrow \quad \quad \nonumber\\
 &\gamma\gamma\quad \: &  \mu^{\pm} \nu_{\mu} \rightarrow e^{\pm}
 \nu_{\mu} \nu_{e} \nu_{\mu} \nonumber
\end{eqnarray}

This mechanism leads to the production of energetic neutrinos, $\nu_{\mu}$ and $\nu_e$, via the $\pi^{\pm}$ decays. The $\pi^{0}$ decay produces energetic photons\footnote{The high energy photons produced by this mechanism could be the ones detected by the \v{C}erenkov telescopes.}. According to the source of the accelerated proton and to the target, several neutrinos sources can be distinguished.

{\bf Diffuse neutrinos sources} are produced by the interaction of the Cosmic Rays with a target which can be the Earth atmosphere, the matter in the galactic plane or the photons of the Cosmic Microwave Background (table \ref{tab:diffuse}). Atmospheric neutrinos is the only source which has been observed so-far. The production of galactic  and cosmologic neutrinos is guaranteed (the source and the targets exist !). However, even if there is a large uncertainty in the computations, all the predictions lead to very low fluxes and it will be difficult to observe them. 
 
 {\bf Point-like neutrinos sources} are obviously more attractive. The observation and the identification of a source will lead to a major progress in the knowledge of the origin of the most energetic Cosmic Rays. Except for the supernova remnants in which protons are accelerated in the expanding shell by the Fermi mechanism, all the sources involve a massive object, a pulsar or more efficiently a black hole. The presence of a very strong magnetic field together with an important plasma flow leads to the production of high energy protons by stochastic acceleration. The target can be very diversified: the surrounding matter for the young supernova remnants, a companion star for the X-ray binary systems, ... (table \ref{tab:pointlike}). A particularly interesting potential source is the AGNs. In their core a massive back hole (typically 10$^8$ M$_\odot$) accelerates protons at ultra high energy. Their interaction with the radiation field and/or the matter surrounding the black hole or in the jets could produce neutrinos. In any case, it must be pointed out that the existence of such point-like sources is still very speculative and is the subject of intense discussions between theorists (see for example \cite{ref2}, \cite{ref3} and \cite{ref4}). In any case, the best way to conclude the debate is to observe or not these neutrinos.  
  
{\bf Gamma ray bursts} have a very peculiar place in this bestiary. They have been observed for many years and are currently observed at a rate of about one event per day. The bursts could be generated by highly relativistic shocks, within a concentrated fireball, which would give rise to the production of neutrinos during a very short time interval \cite{ref5}. The expected neutrinos fluxes are very low (their observation could be possible by making use of correlations with the direction and the timing of the signal seen by a gamma ray detector).

\begin{table}[h]
\begin{center}
\begin{tabular}{*{2}{l}}
\hline
 {\bf Diffuse sources} & {\bf Target}                  \\
\hline
 Atmospheric neutrinos & Earth atmosphere              \\
\hline
 Galactic    neutrinos & Matter in the galactic plane  \\
\hline
 Cosmologic  neutrinos & Photons of the CMB            \\
\hline
\end{tabular}
\caption{Diffuse neutrinos sources produced by the interaction of the Cosmic Rays with different targets. Only the atmospheric neutrinos have been directly observed so-far. The others diffuse sources are guaranteed, but the fluxes are very low.}
\label{tab:diffuse}
\vspace{9mm}
\begin{tabular}{*{3}{l}}
\hline
 {\bf Point-like sources}& {\bf Source of proton}   & {\bf Target}        \\
\hline
 Young supernova remnant& Plerion or expanding shell& Expanding shell     \\
\hline
 Supernova remnant     & Expanding shell           & Expanding shell     \\
\hline
 X-ray binary system   & Pulsar or black hole      & Accreting matter or  \\
                       &                           & companion star      \\
\hline
                       &                           & Matter surrounding  \\
 AGN    	           & Massive black hole	   & the black hole or     \\
 		           & 			         & ambient radiation field     \\
\hline
\end{tabular}
\caption{Possible point-like neutrinos sources as a function of the origin of the accelerated protons and of the target.}
\label{tab:pointlike}
\end{center}
\end{table}

\newpage

\subsubsection{Non-accelerating mechanisms}

 Neutrinos could be also produced by the annihilation or the decay of elementary heavy particles predicted by some theoretical models in Particle Physics.
 
 One of the most promising extension of the Standard Model in Particle Physics is the Minimal SuperSymmetric Model (MSSM). In the simplest version of this theory the Lightest Supersymmetric Particle (LSP) is the lightest neutralino which is stable and massive (its mass is constrained experimentally above $\sim$50 GeV and theoretically below $\sim$1 TeV). It is a natural candidate to the dark matter which composes most of the matter of the Universe. The LSPs which have been produced just after the Big Bang, move in the Galaxy halo at few hundreds km/s. They can lose energy by elastic scattering on nuclei of the Earth and the Sun matter. They can be trapped by the gravitational field of the Earth or the Sun and can be concentrated in their centres where they would annihilate\footnote{Neutralinos are Majorama particles: they are their own anti-particle.} and would produce neutrinos.
 
 A more exotic candidate for the neutrinos production is the decay of very massive particles which could have been abundantly produced after the Big Bang. The decay of monopoles or cosmic strings would release a huge energy which could produce high energy neutrinos.  

\subsubsection{Neutrinos fluxes}

\begin{figure}
 \begin{minipage}[t]{.495\linewidth}
  \centering
 \epsfig{figure=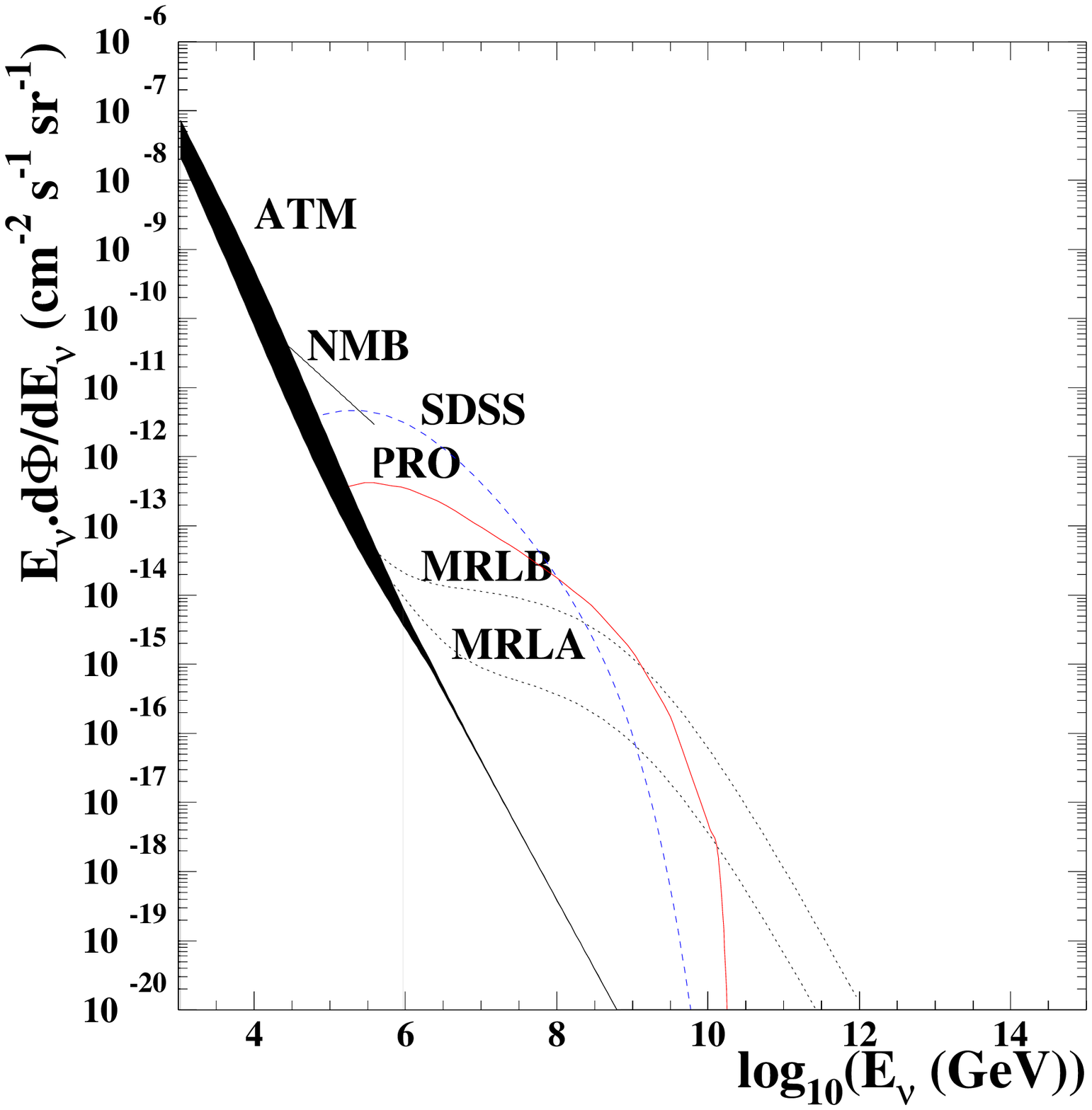, width=.95\linewidth}
 \caption{Expected diffuse neutrinos fluxes from AGNs models. For the models NMB \cite{ref6} and SDSS \cite{ref7}, the neutrinos are produced in the vicinity of the massive black hole. For the models PRO \cite{ref8}, MRLA and MRLB \cite{ref9}, they are produced in the jets. The irreducible background due to the atmospheric neutrinos ATM  \cite{ref10} is also indicated.}
 \label{fig:fluxagn}
\end{minipage} 
\hfill 
 \begin{minipage}[t]{.495\linewidth}
  \centering
 \epsfig{figure=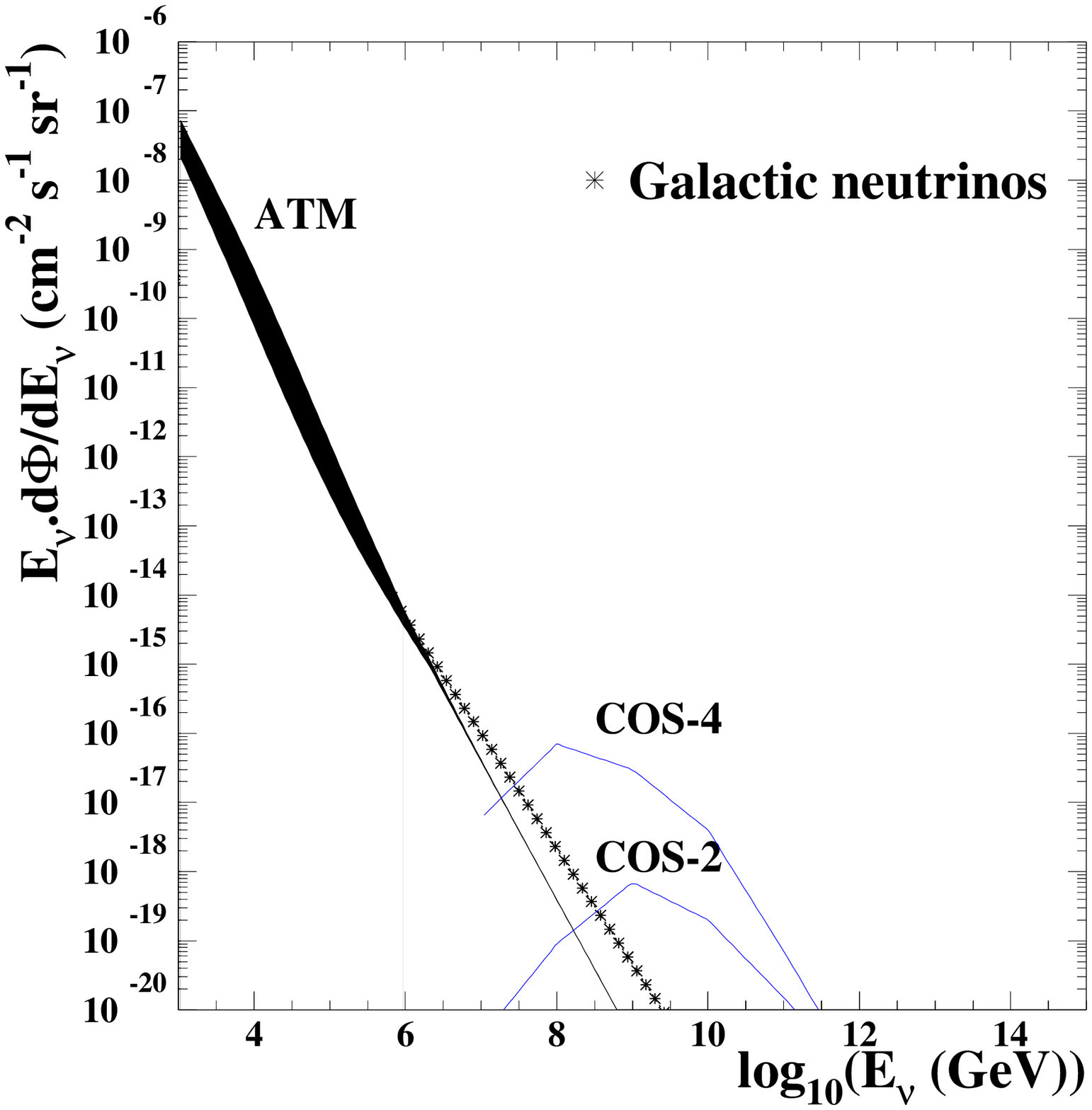, width=.95\linewidth}
 \caption{Expected diffuse flux from the galactic neutrinos \cite{ref11}. Two models of cosmologic neutrinos using two different normalisations are also displayed (COS-2 and COS-4) \cite{ref11}.  The irreducible background due to the atmospheric neutrinos ATM  \cite{ref10} is indicated.}
\label{fig:fluxatm}
\end{minipage} 
\end{figure}

  The figures \ref{fig:fluxagn} and \ref{fig:fluxatm} show the expected neutrinos fluxes from a selection of different sources. Two important remarks can be extracted from these figures:
\begin{itemize}
  \item Neutrino fluxes are very low and a large detector will be necessary: about \hbox{1 km$^2$.}
   \item If we want to observe astronomical diffuse sources, the detector must be able to observe neutrinos with an energy larger than about 10 TeV. Below this limit, atmospheric neutrinos dominate all the other potential sources.  
\end{itemize} 

\newpage
 
\subsubsection{Neutrino oscillation}

  Atmospheric neutrinos are an irreducible background for the study of high energy cosmic neutrinos. However, at lower energy (typically between 5 and 60 GeV), they can be a very interesting signal for the neutrino oscillation study.

  The SuperKamiokande experiment has announced recently the observation of a $\nu_\mu$ deficit \cite{ref1} which is interpreted as the evidence for $\nu_\mu$ mixing with either $\nu_\tau$ or a sterile neutrino. The mixing angle has been found to be $sin^2 (2 \theta) > 0.82$ and the most likely value of the square mass difference is $\Delta m^2 = 2.2 \cdot 10^{-3}$\,eV$^2$ with a confidence interval between $5 \cdot 10^{-4}$ and $6 \cdot 10^{-3}$ eV$^2$.  
  
  A possible method to cross check this result is to measure the energy spectrum of the atmospheric neutrinos. With the hypothesis of two-neutrino mixing, the oscillation probability is given by:
\begin{equation}
 P = \sin^2 (2 \theta) \sin^2 \left( 1.27 \frac{L}{E} 
\Delta m^2 \right)   \label{proba_oscill}
\end{equation}
where $\theta$ is the mixing angle, $L$ is the distance travelled by the neutrino in km, $E$ is the neutrino energy in GeV and $\Delta m^2$ is the difference of the square of the masses in eV$^2$. 

If only the upward going neutrinos are studied, the baseline $L$ is constant (the Earth diameter, $\sim$12740 km). Moreover, if the SuperKamiokande result is correct, the first maximal neutrino flux suppression will be observed at an energy given by: 
\[ E = 10 \times \left( \frac{\Delta m^2}{10^{-3}{\mathrm{eV}}^2} \right) 
    {\mathrm{GeV}}  \]
 
Other values of the series can be calculated by dividing this first value by an odd integer number. Therefore, an important dip in the energy spectrum will be visible for the upward going neutrinos in the energy range of 5 to 60 GeV.

\section{The detection of neutrinos}

\subsection{Principle}

A cheap and efficient method to observe a muonic neutrino is to detect the muon produced from its interaction with the matter surrounding the detector. When passing through water or ice, the muon emits \v{C}erenkov light which is detected by a three-dimensional matrix of photomultiplier tubes. The measurement of the arrival time of the \v{C}erenkov light on the photomutipliers allows to reconstruct the muon direction, the amount of light giving an estimate of its energy.  

Moreover the $\nu$N cross-section and the muon range increase with the energy: the probability to detect a muon coming from a neutrino increases with the neutrino energy. The most energetic neutrinos are therefore enhanced, improving the detector performance at high energy.

\subsection{Background}

 The observation of astronomical sources is subject to two sources of background: the neutrinos and the muons which are copiously produced by the Cosmic Rays interaction with the Earth atmosphere. 
 
 Atmospheric neutrinos are an irreducible source of background for the study of the cosmic neutrinos (how to distinguish a neutrino produced in the Earth atmosphere from one produced inside an AGN ?). However, as it is visible on the figure \ref{fig:fluxagn}, they follow a E$^{-3}$ law and their flux becomes less important than the one of the signal beyond 10-100 TeV. 
 
\begin{figure}
 \begin{minipage}[t]{.495\linewidth}
  \centering
  \epsfig{figure=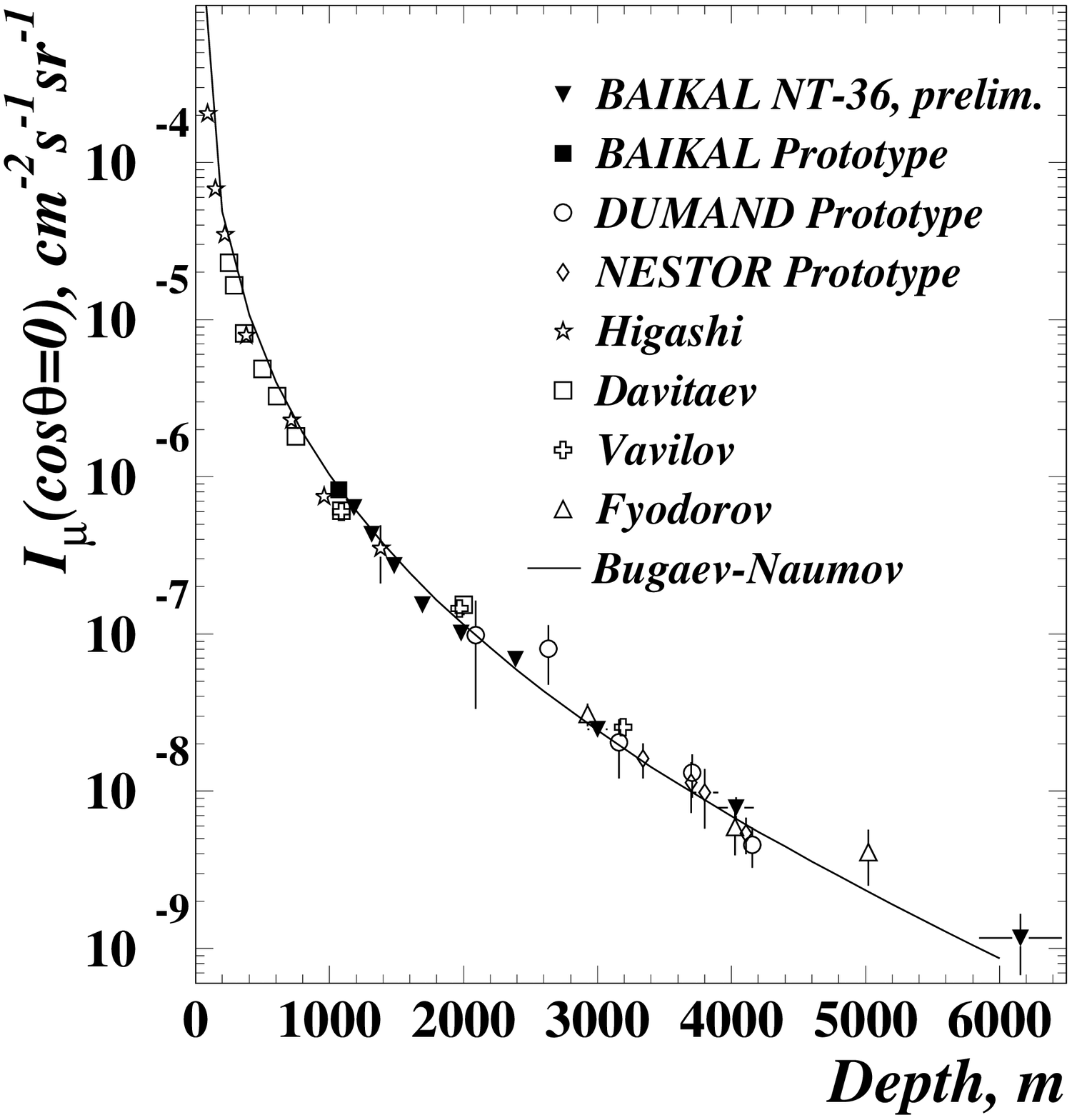, width=.95\linewidth}
  \caption{Flux of downward going muons as a function of the detector depth in water equivalent.}
  \label{fig:muon1}
 \end{minipage} 
\hfill 
 \begin{minipage}[t]{.495\linewidth}
  \centering
  \epsfig{figure=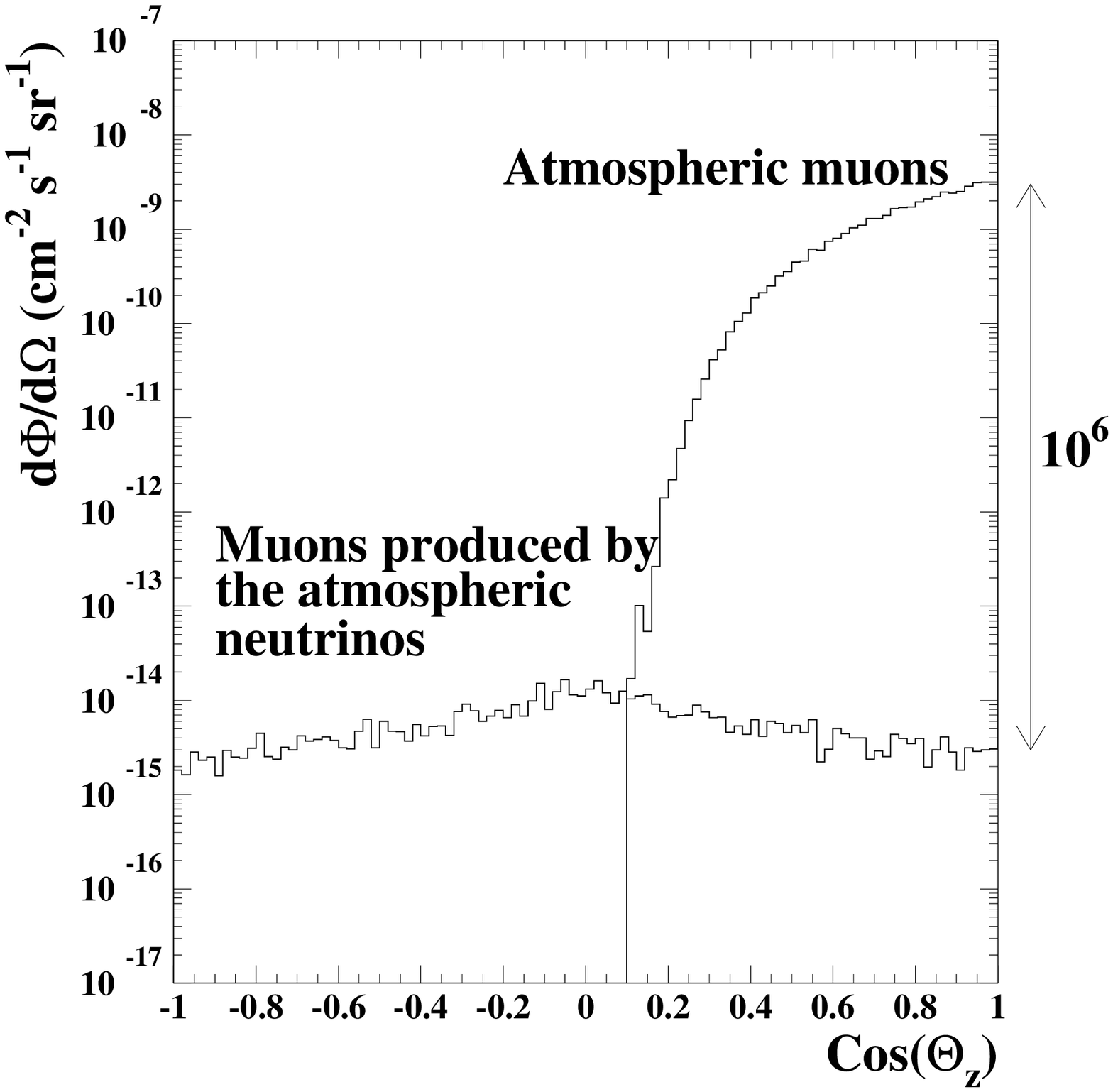, width=.95\linewidth}
  \caption{Muons flux as a function of the zenith angle ($\Theta_z=0^\circ$ corresponds to downward going muons), at a depth of 2400 m water equivalent.}
  \label{fig:muon2}
 \end{minipage} 
\end{figure}

 The protection against the atmospheric muons is a more delicate task and have an important impact on the detector design:
 \begin{itemize}
  \item A very important shielding is necessary (figure \ref{fig:muon1}). The muons flux can be decreased by more than 4 orders of magnitude by installing the detector at 2400 meters deep. 
  \item Nevertheless this protection is not sufficient: muons flux dominates atmospheric neutrinos flux by more than 6 orders of magnitude at small zenithal angle at 2400 meters deep (figure \ref{fig:muon2}). Even at 4000 meters deep, it overcomes the atmospheric neutrino flux by several orders of magnitude and is too important. However, muons are easily stopped by the Earth matter and no upward going muons are observed: to suppress this background, upward going particles must be detected.
\end{itemize} 
  
\subsection{Constraints on the detector}

From the following remarks, the most important constraints on the detector design can be summarised: 
\begin{itemize}
  \item Signal fluxes are very low: the detector surface must be important, about \hbox{1 km$^2$.} However a 0.1 km$^2$ surface should be sufficient to start the study of the cosmic neutrinos.
  \item Due to the background, the detector must:
  \begin{itemize}
    \item Be well shielded.
    \item Observe upward going muons to remove the atmospheric muons.
    \item Have a good energy resolution to reconstruct the energy spectra.
    \item Have a good angular resolution to isolate properly a point-like source and optimise the signal to noise ratio (as discussed in the section {\it 7.3}, the mean number of background events is directly proportional to the pixel size which in turn depends of the angular precision). 
  \end{itemize}
\end{itemize}

\section{The ANTARES collaboration}

The ANTARES\footnote{ANTARES stands for Astronomy with a Neutrino Telescope and Abyss environmental RESearch.} collaboration \cite{ref12} aims at building a large undersea high energy neutrino telescope. This collaboration, which started in the Spring of 1996, includes astronomers, astrophysicists, oceanographers and particle physicists from France, the Netherlands, Russia, Spain and United Kingdom. The detector should be sufficiently multi-purpose to be able to observe astrophysical sources and to study particle physics topics. 

The construction of a km$^2$ undersea detector is a very complex challenge which must be reached by steps. In this objective the ANTARES collaboration aims at building a 0.1 km$^2$ detector which should be able to observe cosmic neutrinos and to demonstrate the feasibility of such a detector. For that the collaboration has defined a R\&D program which can be divided into three main parts:
\begin{enumerate}
  \item Learning of the knowhow to install and to operate strings equipped of photomutipliers and connected to the shore. 
  \item Measurements of the environmental parameters in deep sea to characterise and identify the sites suitable to the installation of a neutrino detector.
  \item Computer simulations to optimise the detector geometry and the analysis algorithms.
\end{enumerate}

The km$^2$ detector site has not been chosen yet. It will be guided by the properties of the site itself but also of the closeness to the shore, the meteorological conditions, the support available onshore and the boat disponibilities. During the R\&D program a convenient site in the Mediterranean sea situated at about 20 miles of the Porquerolles Islands (south of France) and at \hbox{2400 m} deep has been used. The measured properties of this site (discussed hereafter) have shown no major drawback and it will be used for the next step of the project, the \hbox{0.1 km$^2$} detector.
 
 The different points of the R\&D program will be now discussed in more details.

\section{Development of the strings}

\subsection{Detector layout}

  The final detector will have a large number of instrumented vertical strings. These ones will be installed on the sea bottom to form a tridimensional matrix of photomultipliers (the distance between the strings will be about 80 meters). The whole detector will be connected to the shore by an electro-optical cable which ensures the data transmission and the power supply. 
  
   Each string is about 400 meters long and is equipped with about 100 photomultipliers grouped by floors, with an inter-floor distance between 8 m and 15 m. To facilitate their deployment, the strings are flexible. Therefore an accurate positioning system is necessary to monitor their shape during and after the deployment (deep sea currents modify the shape). Some other instruments are also necessary to monitor the properties of the sea water (water optical properties, monitoring of the glass transparency, ...), to calibrate in time and in energy the detector, ... 
 
 The conception of such a detector is not an easy task. To succeed in this operation, the development has been divided in several parts which are pursued in parallel. We will now give more details in the three most important studies: the development of the Optical Modules (``the eyes'' of the detector), the design and the deployment of a string prototype and the tests of connection in deep sea.

\subsection{The Optical Module}

\begin{figure}
\centering
\epsfig{file=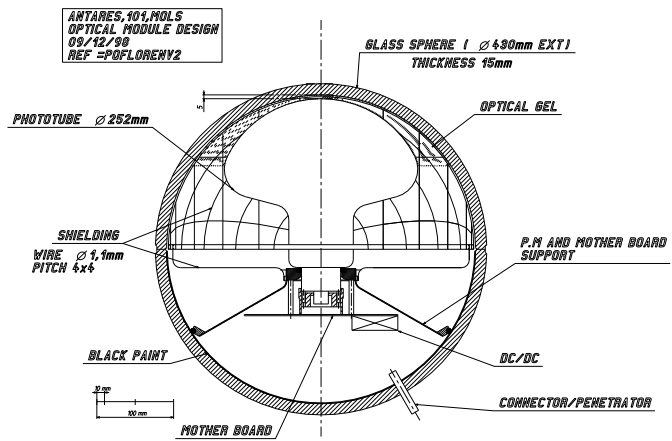,width=9cm}
\caption{Sketch of an Optical Module. A large hemispherical photomultiplier (diameter greater than about 8 inches) is protected from the ambient pressure by a  pressure-resistant glass sphere. The outer diameter of this sphere is 17 inches (about 42.5 cm).}
\label{fig:om}
\end{figure}

  Due to the ambient pressure at 2400 meter deep, photomultipliers must be housed inside pressure-resistant glass spheres (figure \ref{fig:om}). In these so-called Optical Modules the photomultiplier is embedded in silicon gel to ensure a good optical coupling and to maintain it mechanically. To optimise the light detection, hemispherical photomultipliers with a photocathode surface larger than 8 inches and with a very good time resolution are used. A high permittivity alloy cage surrounds it, shielding it against the Earth magnetic field. A DC/DC converter supplying power to the photomultiplier is also included in the  sphere. Signal outputs, high voltage monitoring, ... are transmitted through water and pressure-resistant connectors to the outside world.

 Digital transmission will be used to transmit data to the shore. The digitisation of the photomultiplier signal will be performed at the Optical Module level. An original design for an Application Specific Integrated Circuit (ASIC) has been developed by the ANTARES collaboration. It samples at a frequency up to 1 GHz and memorises analog pulses in an array of MOS capacitors. Data corresponding to the very frequent single photoelectron pulses are not numerized, only a minimal information (charge and arrival time) is transmitted to the shore. 
 
\subsection{Test of the string prototype}

\begin{figure}
\centering
\epsfig{figure=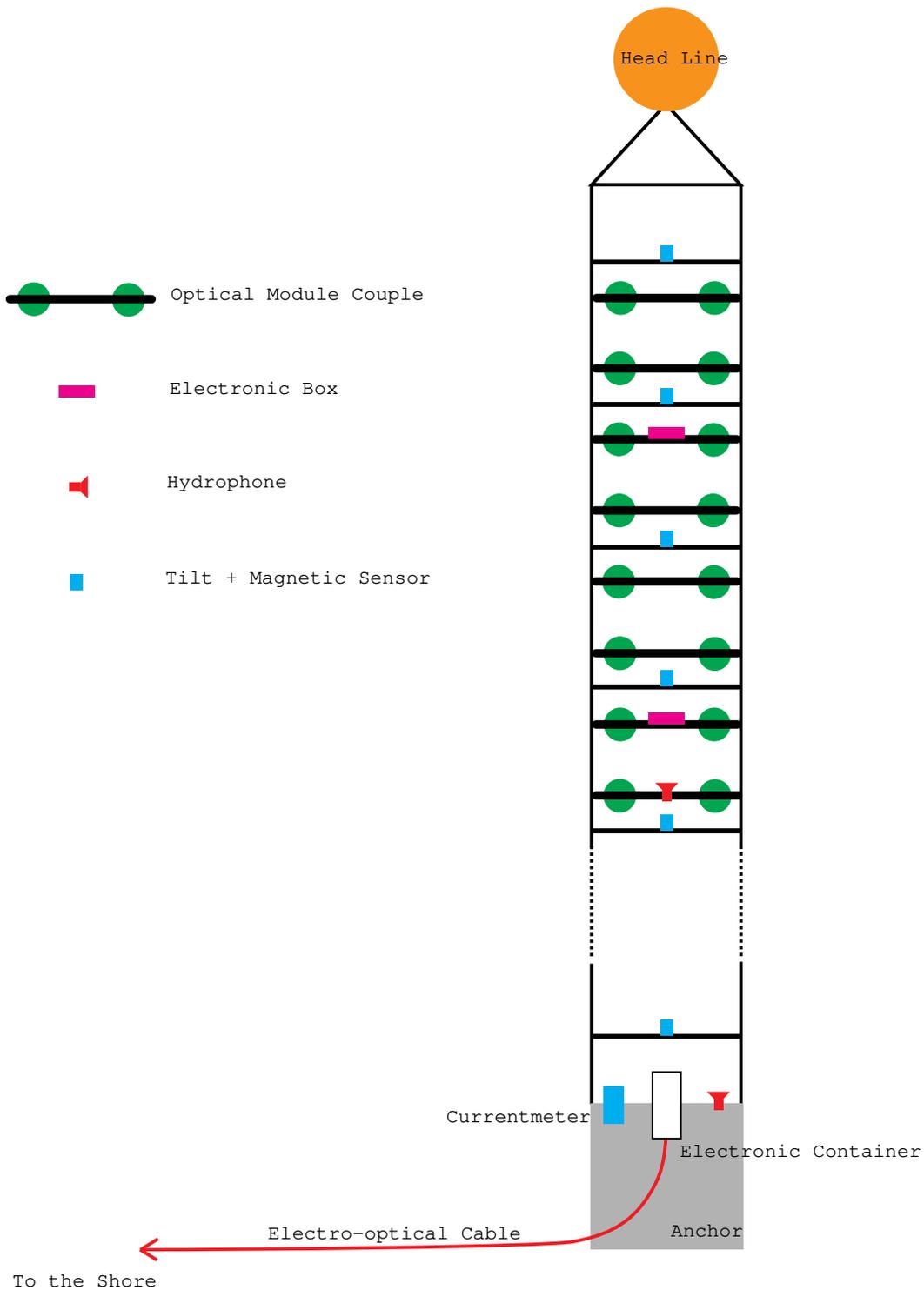, width=14cm}
\caption{Schematic view of the first string prototype. An Optical Module is composed of a photomultiplier housed inside a pressure resistant glass sphere. The hydrophons, compass and tiltmeters are used to monitor the string shape and to measure very precisely the position of each Optical Module.}
\label{fig:line5}
\end{figure}

  A first string prototype has been developped to learn the knowhow to build, deploy and operate such a structure. A sketch of this string is shown on the figure \ref{fig:line5}.

  Even if this string prototype will be equipped of only 8 photomultipliers, all the elements necessary for the good running of the detector have been installed: slow control, energy calibration, time calibration, positioning (the system composed of hydrophon, compass and tiltmeter allows the monitoring of the position of each photomultipliers with a precision better than 10 cm), ... To decrease the influence of the optical background, the photomultipliers are clustered by two and can be put into coincidence. One should note that due to the  development schedule of the digital transmission described previously, a first version of the signals readout and transmission using a simple on the shelf analog electronics is being used. However the digital transmission will be used for the next strings.

  To test the mechanical behaviour of this string in deep sea, this one has been successfully deployed in July-August 1998. In the Spring 1999, it will be equipped of 8 photomultipliers, deployed and connected to the shore via an electro-optical cable. Data should be taken for several months. 

 It must be emphasized that the design of the strings foreseen for the next step of the project will be slightly different. For example, computer simulations have shown that the Optical Modules must be clustered by three to optimize the detector performances  (they are clustered by two for this prototype).

\subsection{Tests of connection in deep sea}

  As already pointed out, an electro-optical cable will be used to connect the detector to the shore. It would not be reasonable in terms of cost and efficiency to have one cable per string. Therefore, each string will be connected to a junction box which in turn will be connected to the shore via the electro-optical cable.
  
  To connect the strings to the junction box, it is foreseen to use a submarine. This one will be furnished by the IFREMER \cite{ref13} which participates to the ANTARES collaboration. This institute owns in particular two deep sea submarines, the NAUTILE (up to 6000 m deep) and the CYANA (up to 3000 m deep) which can be used for this operation. A first test of deep sea connection has been successfully performed in December 1998 with the NAUTILE. These tests will be pursued in the next years.

\section{Measurements of the environmental parameters}

\subsection{Introduction}

  The detector geometry and its performances are directly linked to some environmental parameters:
\begin{itemize}
  \item {\bf The optical background}. Due to the $^{40}$K decay and to the life activity in sea water (the bioluminescence), the optical background has important constraints on the trigger logics and the electronics as well as the mechanical layout of the photomultipliers.
  \item {\bf The effect of the biofouling and the sedimentation on the glass transparency}. Due to the ambient pressure at 2400 m deep, photomutipliers are housed inside a pressure resistant glass sphere. Biofouling and sedimentation deposition can diminish the glass transparency.
  \item {\bf The water optical properties}. The absorption and the scattering length influence the detector design (distance between strings and between Optical Modules on the same string). 
\end{itemize}
  
\begin{figure}
\centering
\psfig{figure=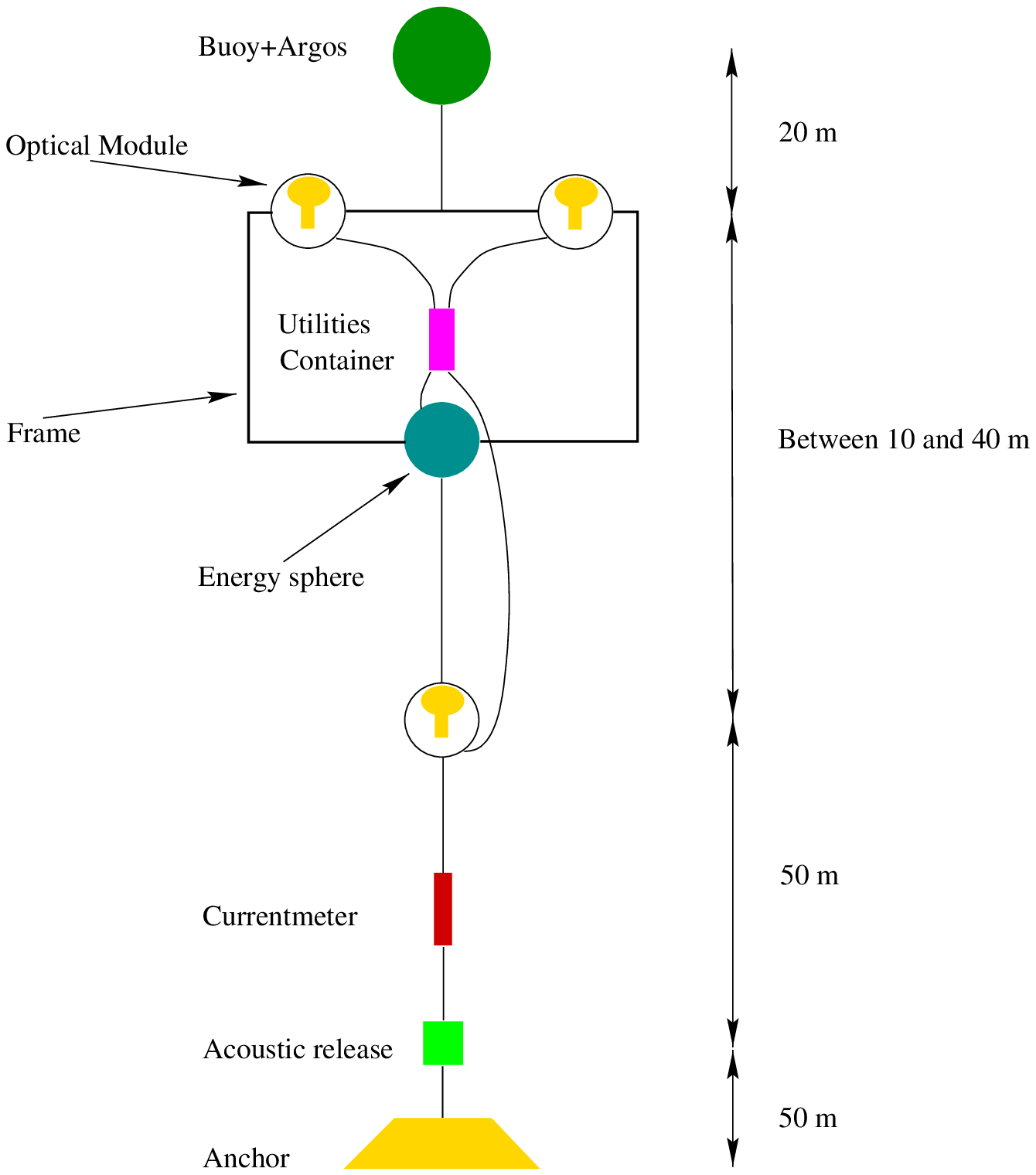, width=9cm}
\caption{Schematic view of the string used to measure the optical background. The string is maintained vertically by the buoy. When the measurement is achieved, the acoustic releases are activated and the string floats back to the surface where a boat can recover it.}
\label{fig:line1}
\end{figure}

 To measure these parameters, the ANTARES collaboration has built and deployed several autonomous strings allowing long term measurements (these environmental parameters are obviously local and seasonal). Each set-up is mounted on a mooring line which is held vertically by a buoy (see for example the figure \ref{fig:line1}). It is placed at about 100 meters from the sea bottom which might be muddy. All the strings are immersed from ships belonging to the French Insitut National des Sciences de l'Univers (INSU-CNRS).

 More than 50 trips with the INSU-CNRS ships have been already performed. Most of the measurements have been done at a depth of approximatively 2400 m, in the current ANTARES site.

\subsection{Optical background}
  
  To measure the spatial extension of the optical background, the string is equipped of three Optical Modules identical to the ones used for the string prototype (figure \ref{fig:line1}): two Optical Modules are 0.5 to 1.5 m apart, and the other one is 10 to 40 m away. A current-meter is also installed to analyse the correlation between the sea current and optical background. 

\begin{figure}
 \begin{minipage}[t]{.495\linewidth}
  \centering
  \epsfig{figure=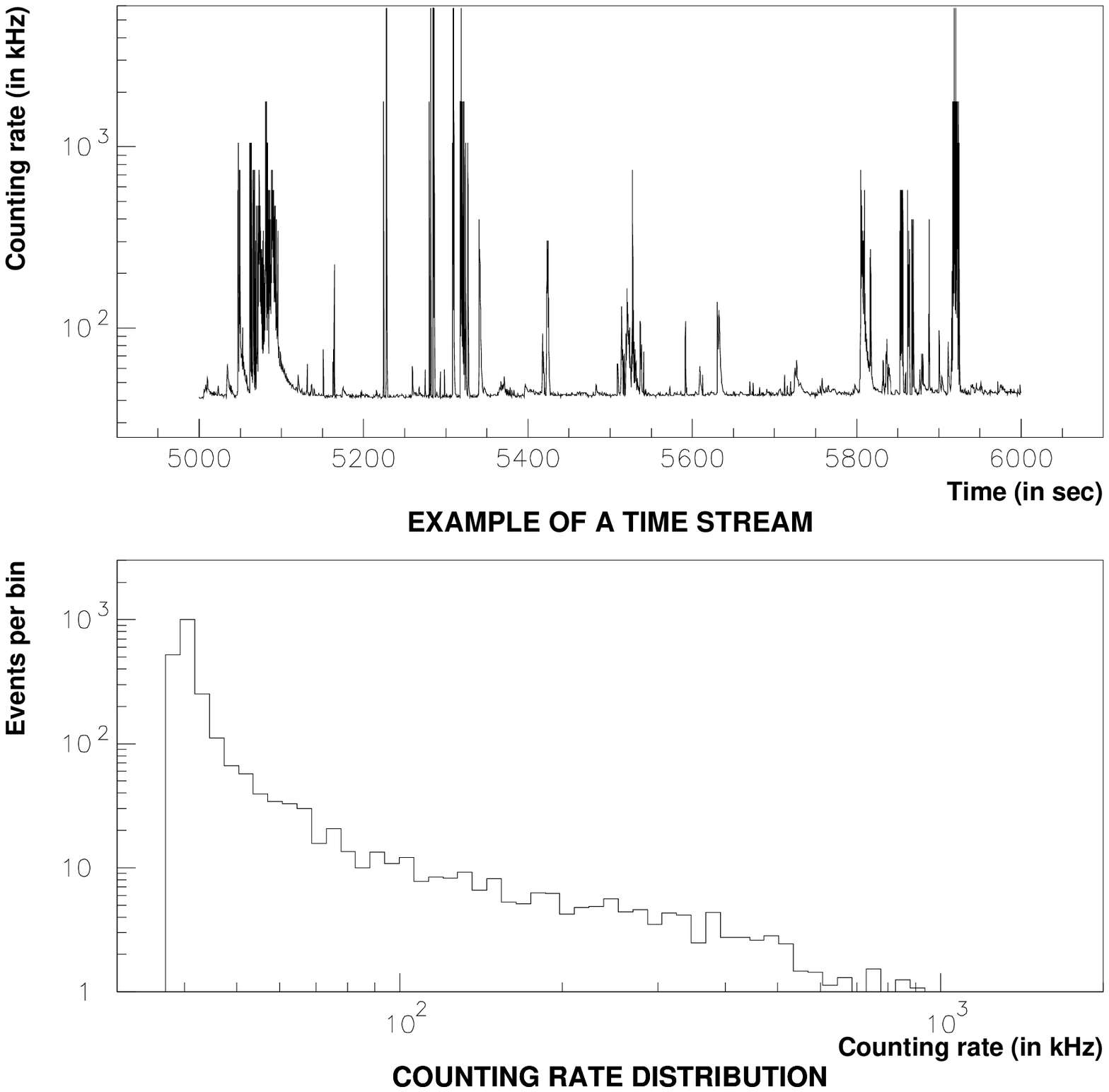, width=.95\linewidth}
  \caption{Counting rate measured by a single 8 inches photomultiplier (upper histogram). The counting rate distribution is shown in the lower histogram. The pulse height threshold has been fixed to 1/3 of the mean amplitude of the single photoelectron signal.}
  \label{fig:k40}
 \end{minipage} 
\hfill 
 \begin{minipage}[t]{.495\linewidth}
  \centering
  \epsfig{figure=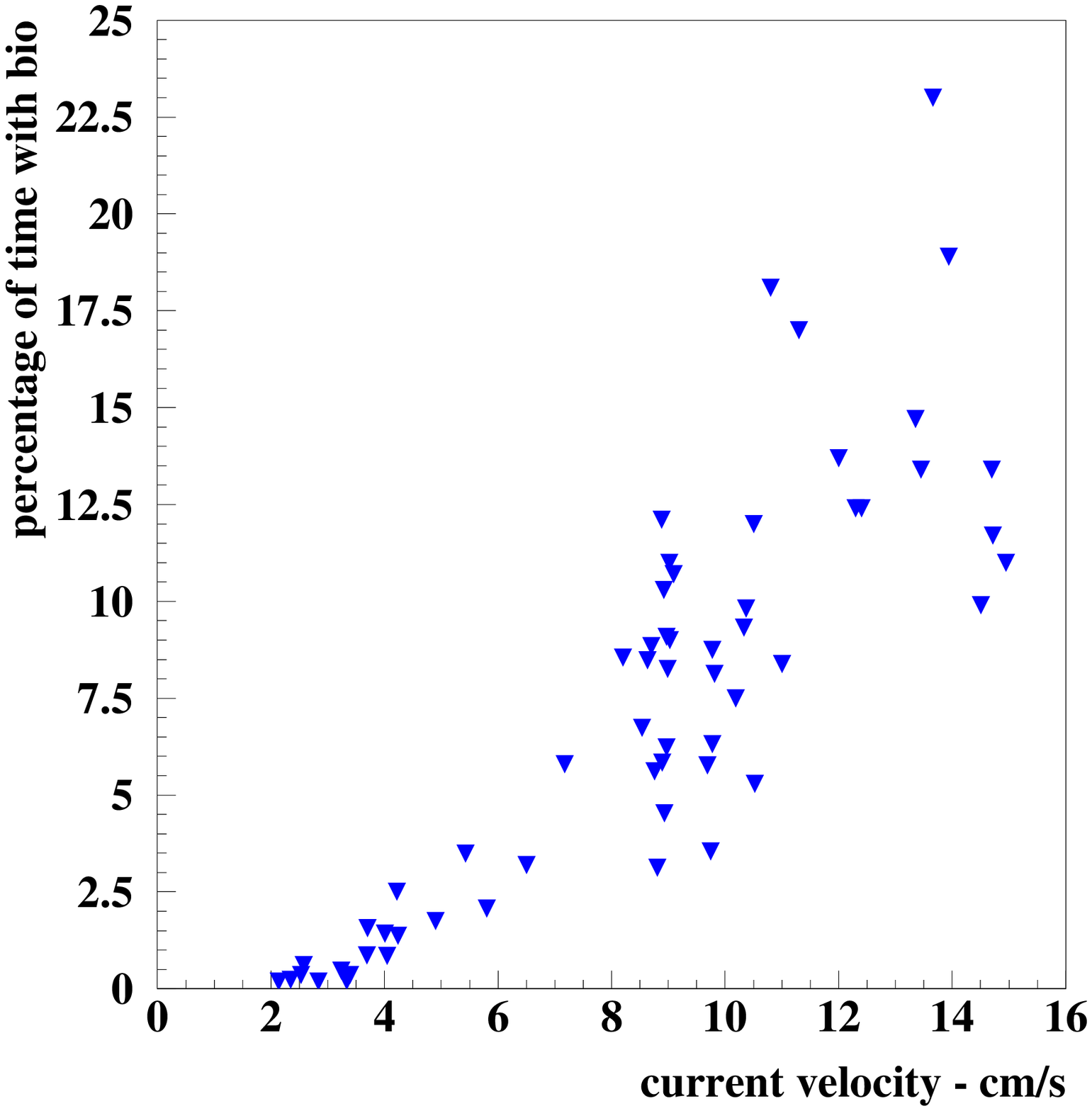, width=.95\linewidth}
  \caption{Percentage of time where the counting rate of the photomultiplier exceeds by 15 \% the baseline, as a function of the deep sea current. A clear correlation between the sea current and the bioluminscence activity is visible.}
  \label{fig:current}
 \end{minipage} 
\end{figure}

  The measurement consists in recording the counting rate of the photomultipliers as a function of time and at different signal amplitude. By putting two photomultipliers in coincidence, it is possible to measure the spatial extension of the optical background. The figure \ref{fig:k40} shows the typical counting rate measured by a 8 inches photomultiplier. The baseline at about 40 kHz is due to the $^{40}$K decay and to a constant contribution of the bioluminescence. The bursts are due to sudden bioluminescence activities.   

  From our measurement it appears that:
\begin{itemize} 
  \item By putting two photomultipliers in coincidence, the optical background can be diminished to a reasonable level, lower than 50 Hz for 8 inches photomultipliers.
  \item The dead time due to the bioluminescence bursts is small, between 3 and 4 \%. 
  \item A clear correlation between this bioluminescence activity and the deep sea current has been observed (figure \ref{fig:current}).
\end{itemize} 
  
\newpage

\subsection{Biofouling and sedimentation effect}

\begin{figure}
 \begin{minipage}[t]{.495\linewidth}
  \centering
  \epsfig{figure=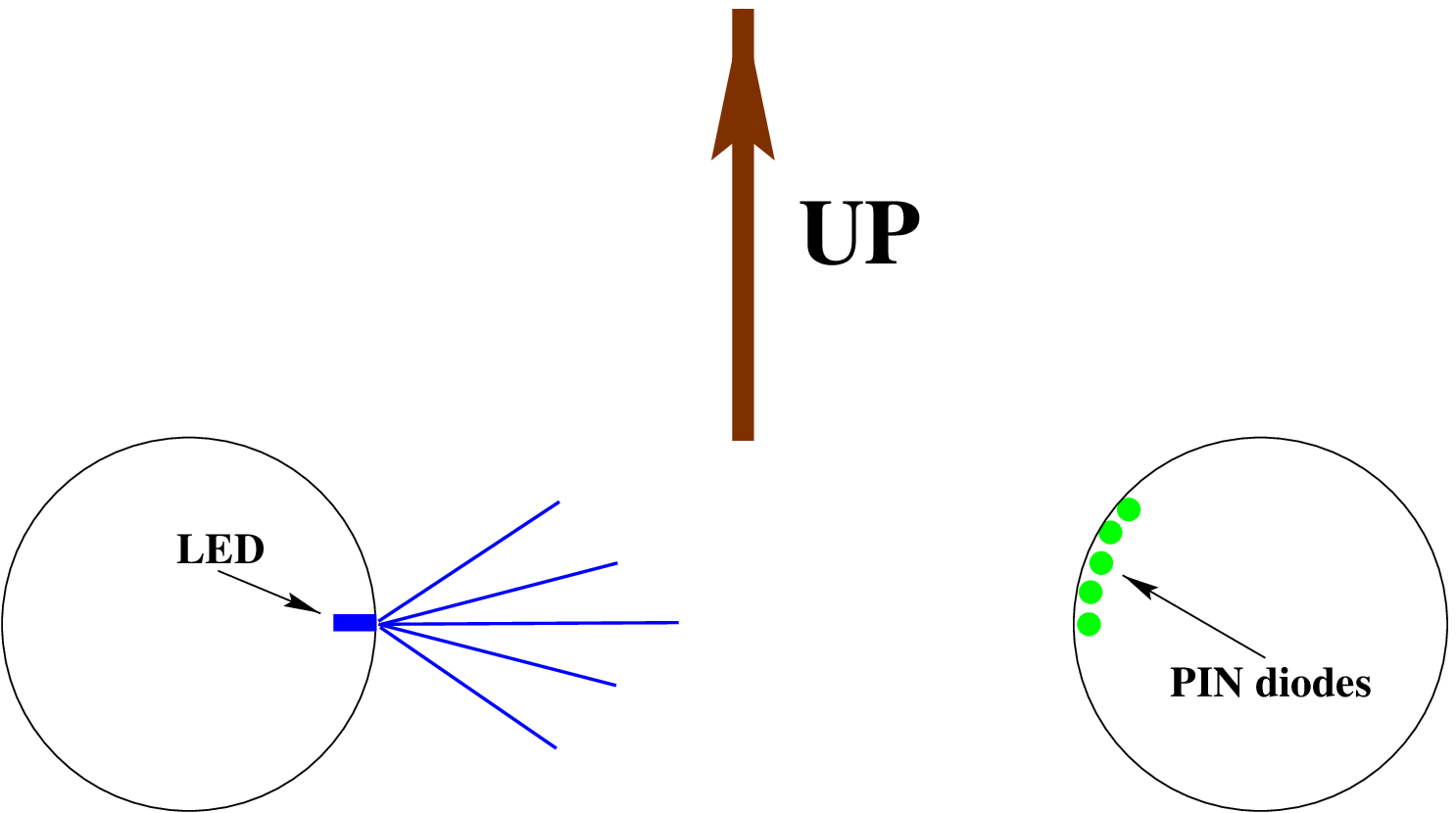, width=.95\linewidth}
  \caption{Positioning of the LED source and of the PIN diodes to monitor the effect of the biofouling and of the sedimentation on the glass transparency.}
  \label{fig:pindiodes}
 \end{minipage} 
\hfill 
 \begin{minipage}[t]{.495\linewidth}
  \centering
  \epsfig{figure=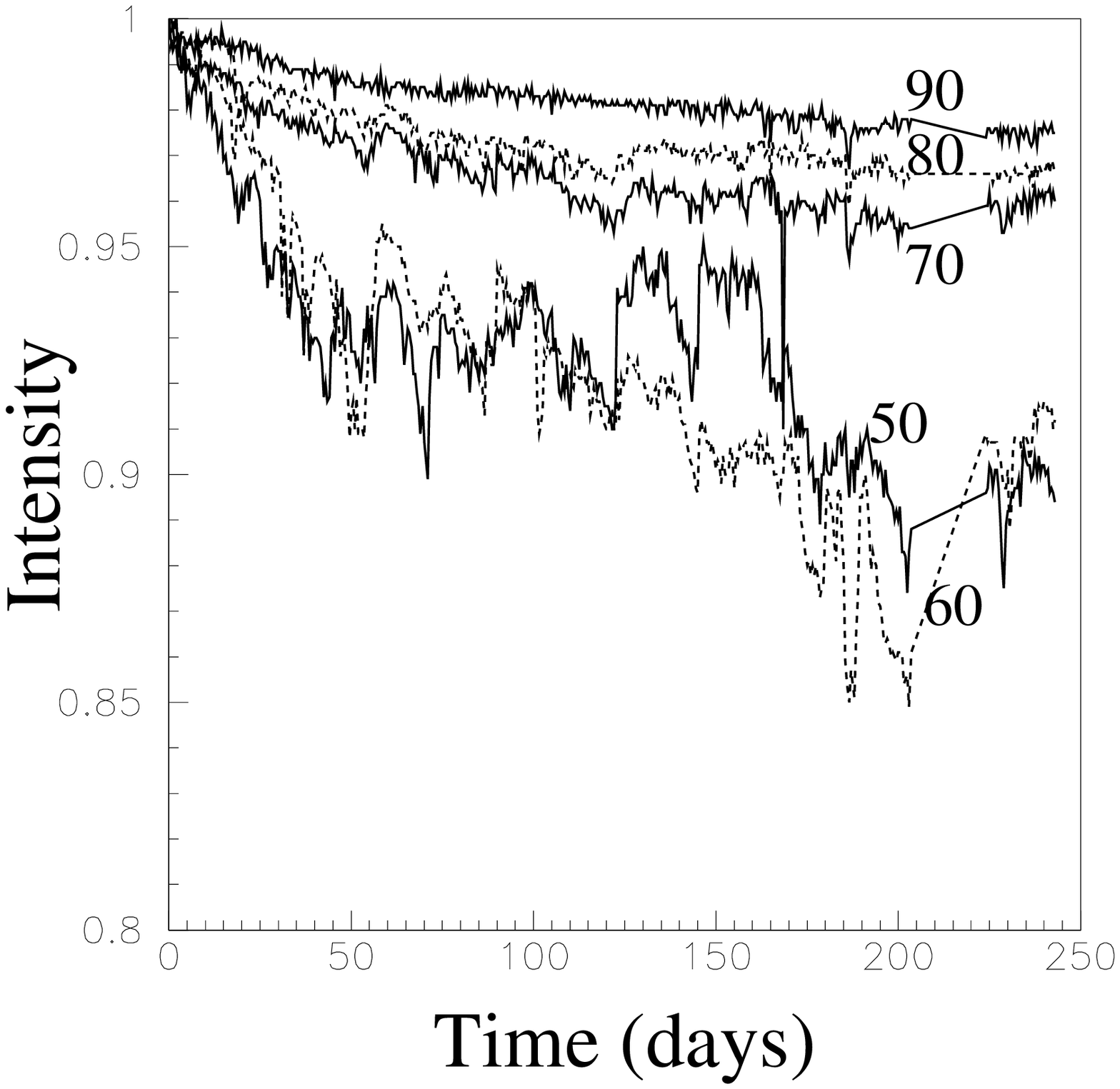, width=.95\linewidth}
  \caption{Light transmission monitored over 240 days (about 8 months). The polar angle of each PIN diodes is indicated (0$^\circ$ corresponds to the top of the sphere and 90$^\circ$ to the equator). The intensities have been normalised to the first measurement.}
  \label{fig:sedim}
 \end{minipage} 
\end{figure}

  To measure the effect of the biofouling and of the sedimentation on the glass transparency, a blue LED source has been placed inside a glass sphere and a set of PIN diodes glued inside a second sphere which was placed one meter apart (figure \ref{fig:pindiodes}). The measurement consists in monitoring the amount of light seen by each PIN diodes as a function of time. To measure a significant effect the set-up must be immersed for several months. Therefore, an acoustic modem has been installed to recover the data from time to time.
  
  The effect on the glass transparency is visible on the figure \ref{fig:sedim}. The influence of the biofouling and of the sedimentation is less important at the equator than at the top of the sphere. If we suppose that the attenuation effect is the same on the two spheres, the attenuation at the equator on one glass sphere doesn't exceed 2.5\%/2=1.25\% in 8 months.  
  
  From this measurement, it has been decided that the photomultipliers must be faced down to be insensitive to the biofouling and to the sedimentation.    
 
\subsection{Water optical properties}

  The water optical properties are very important for the definition of the detector geometry (distance between Optical Modules on the same string and distance between strings) and for its performances. In particular a good scattering length is particularly important to obtain a good angular resolution.

\begin{figure}
 \begin{minipage}[t]{.495\linewidth}
  \centering
  \epsfig{figure=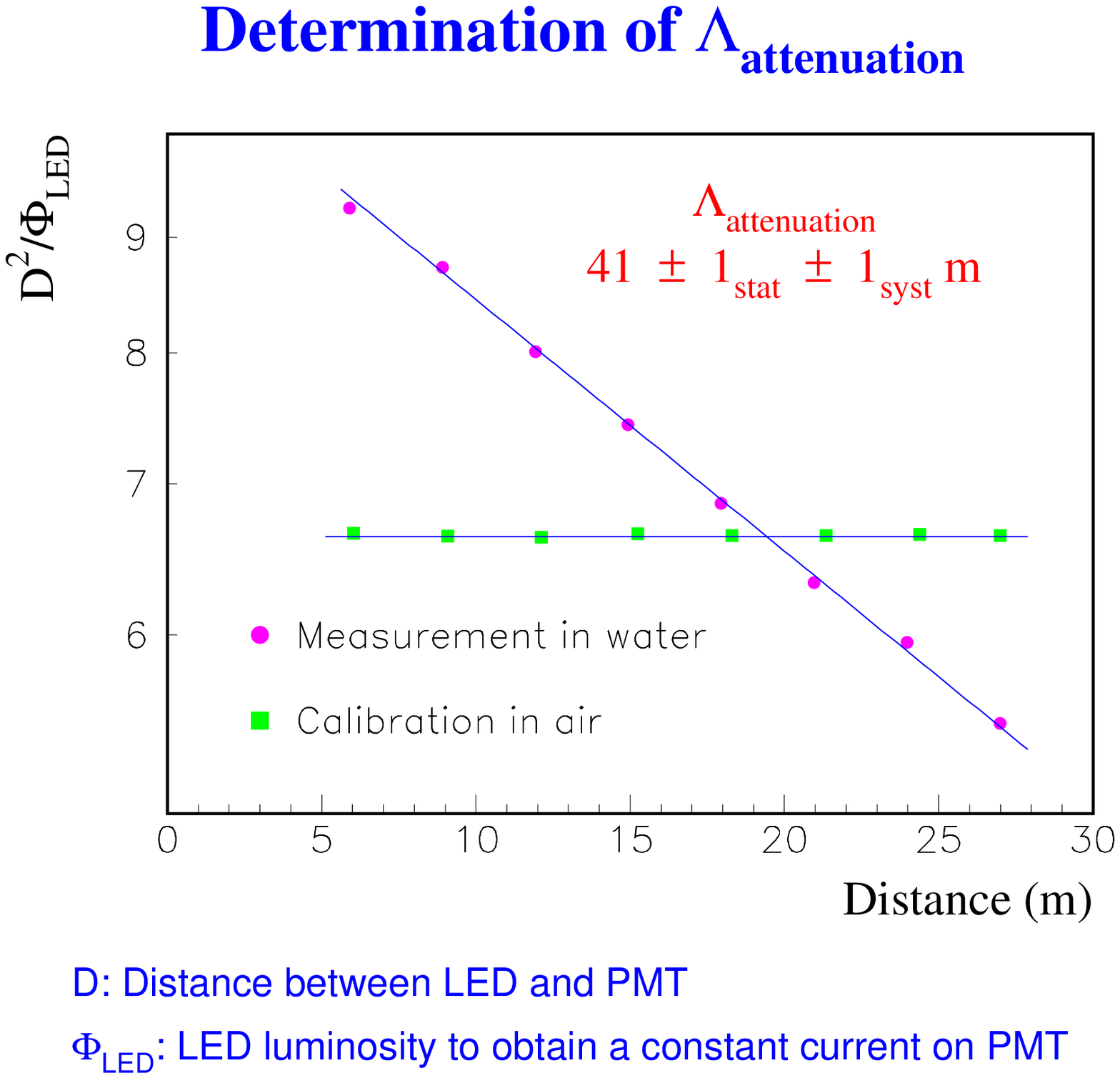, width=.95\linewidth}
  \caption{Attenuation length \hbox{measured by monitoring the amount} of light seen by a photomultiplier ($\Phi_{LED}$) as a function of the distance to the light source ($D$). In air, the amount of light measured follows a standard 1/D$^2$ law and the ratio $D^2$/$\Phi_{LED}$ is constant. In water, the slope is due to the attenuation.}
  \label{fig:att}
 \end{minipage} 
\hfill 
 \begin{minipage}[t]{.495\linewidth}
  \centering
  \epsfig{figure=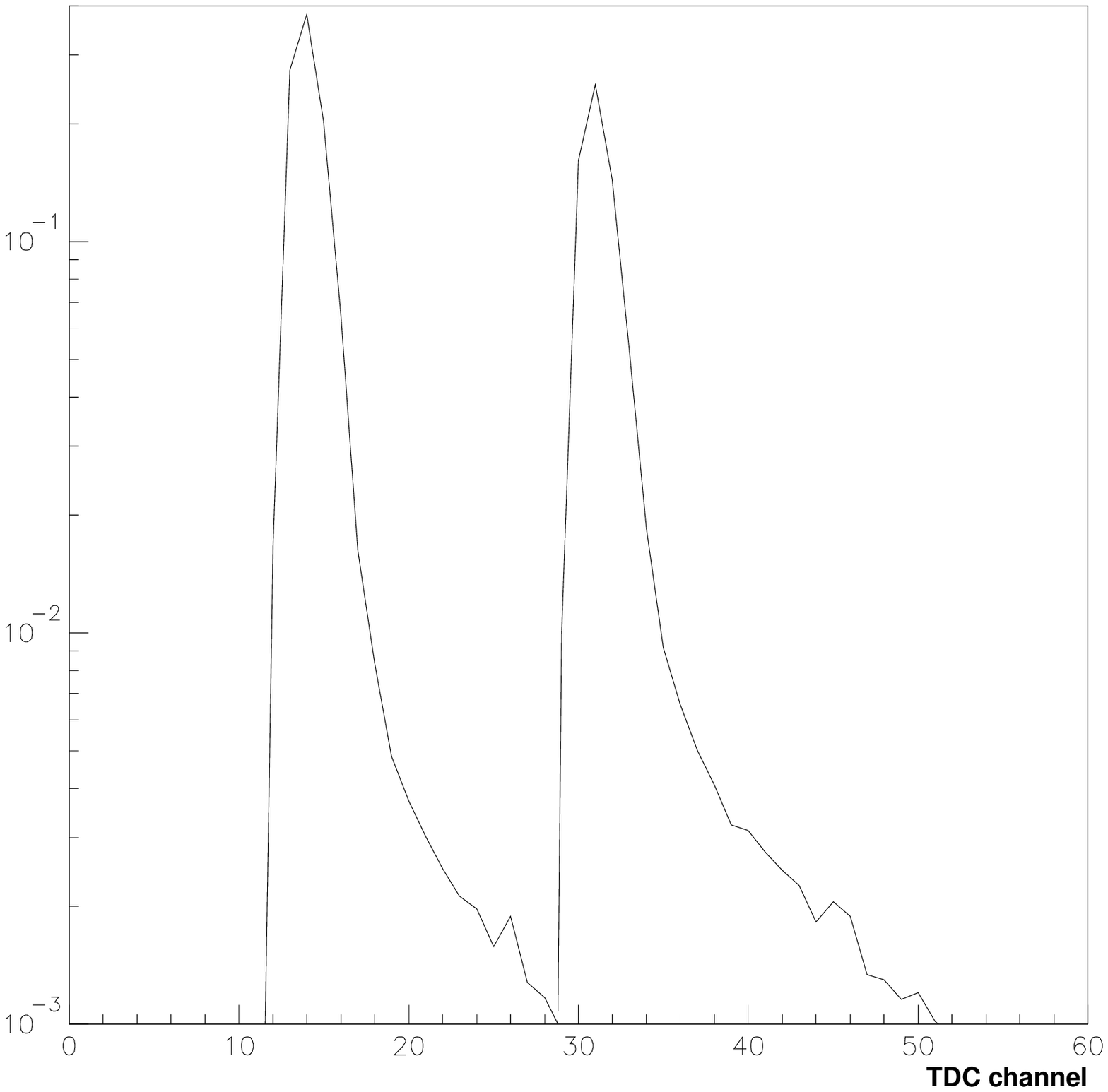, width=.95\linewidth}
  \caption{Time distribution of the arrival time of the photons (1 TDC bin corresponds to 5.5 ns). The distance between the pulsed light source and the photomultiplier is 25 m for the left curve and 44 m for the right curve. The tail is due to the scattered photons. From the shape of these spectra, it is possible to extract the absorption and the scattering length.}
  \label{fig:scatt}
 \end{minipage} 
\end{figure}

  To measure the attenuation, the absorption and the scattering length, two different strings have been conceived:
\begin{itemize}
  \item The first string is equipped of a 32 m long rail used to guide a mobile cart. A light source is placed at the top of the rail and a photomultiplier on the cart. With this set-up the attenuation length can be measured by monitoring the amount of light seen by the photomultiplier as a function of the distance to the source (figure \ref{fig:att}).
  \item A second flexible string has been developped to measure the absorption and the scattering length. The set-up is mainly composed of an isotropic pulsed light source and a 1 inches photomultiplier equipped of a TDC. The distance between then can be either 44 m or 25 m. The measurement consists in monitoring the arrival time of the photons emitted by the pulsed light source: the proportion of scattered photons which are delayed with respect to the direct photons is linked to the scattering length  (figure \ref{fig:scatt}). 
\end{itemize}  
  
 With these two strings, it has been possible to measure in-situ the water optical properties. An attenuation length of 41 $\pm$ 1$_{stat}$  $\pm$ 1$_{syst}$ meters in the blue (460 nm) has been measured and a scattering length greater than 100 meters. From our measurement, it appears that the water optical properties in deep sea are promising.
  
\section{Computer simulations}

  Computer simulations are mandatory to optimise the detector geometry and to estimate its performances.
 
 \subsection{Detector geometry}
  
  The muon track is defined by 5 parameters: one point on the track (2 coordinates), two angles and one reference time. To reconstruct these parameters the arrival time of the photons on the photomultipliers is used. Therefore, only 5 hit photomultipliers are necessary to reconstruct a muon track. 

  The estimation of the muon energy is a more delicate task:
\begin{itemize}
  \item In the energy range suited for the neutrino oscillation study (between 5 and \hbox{60 GeV),} the muon track is contained in the detector. For the events interacting inside the fiducial volume of the detector, a good estimation of the muon energy can be performed by measuring its range. 
  \item At high energy ($>$ TeV), the muon range is greater than the detector size and the previous method can not be used. A possible method is to use the number of detected photons. However the energy loss of an energetic muon  is dominated by catastrophic energy loss from Bremsstrahlung and pair production, and only a crude estimation of the energy will be possible.
\end{itemize}
  
  Considerable efforts have been performed to optimise the geometry of a 0.1 km$^2$ detector ( many details can be found in \cite{ref14}). It appears that strings must be absolutely arranged in a way to avoid symmetries and the photomultipliers should be slanted at 45$^\circ$ below the horizon (with such an orientation, the detector response over the lower hemisphere is assured). The distance between strings and between Optical Modules on the same strings is not yet clearly defined. For the study of astronomical objects, the most important point is the surface of the detector due to the expected very low fluxes. In this case, the distance between strings would be about 60 meters and the distance between the Optical Modules about 16 meters. At the opposite, for the study of the neutrino oscillation, a better sampling of the signal is necessary to measure accurately the muon range. 

  \subsection{Performances of a 0.1 km$^2$ detector}

  The results given hereafter have been obtained with a geometry optimised for the study of the astronomical objects. The detector is composed of 1000 photomultipliers arranged over 15 strings, the distance between strings is 80 meters and the distance between Optical Modules 16 meters. 
     
\begin{figure}
\centering
\epsfig{figure=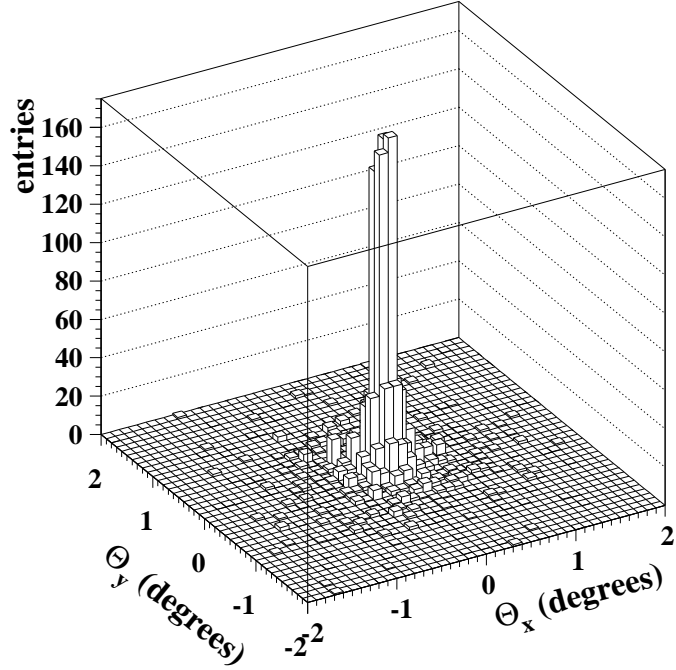, width=10cm}
\caption{Angles between the neutrino and the reconstructed muon as expected with a 0.1 km$^2$ detector. Half of the events have an angular error better than 0.2$^\circ$. The neutrino spectrum which has been simulated follows a E$^{-2}$ law.}
\label{fig:resolangl}
\end{figure}

   With such a detector, a very good angular resolution has be obtained: half of the events have an angular error better than 0.2$^\circ$ (figure \ref{fig:resolangl}). This angular resolution is dominated by the error on the reconstruction: the angular spread due to the physical process is negligible above few TeV. Moreover, even if this angular precision seems poor according to the astronomical telescopes and radio-telescopes (0.2$^\circ$ is the Moon or the Sun radius !), such a precision is the same as the one obtained with the current \v{C}erenkov telescopes which observe TeV photons.

\begin{figure}
\centering
\epsfig{figure=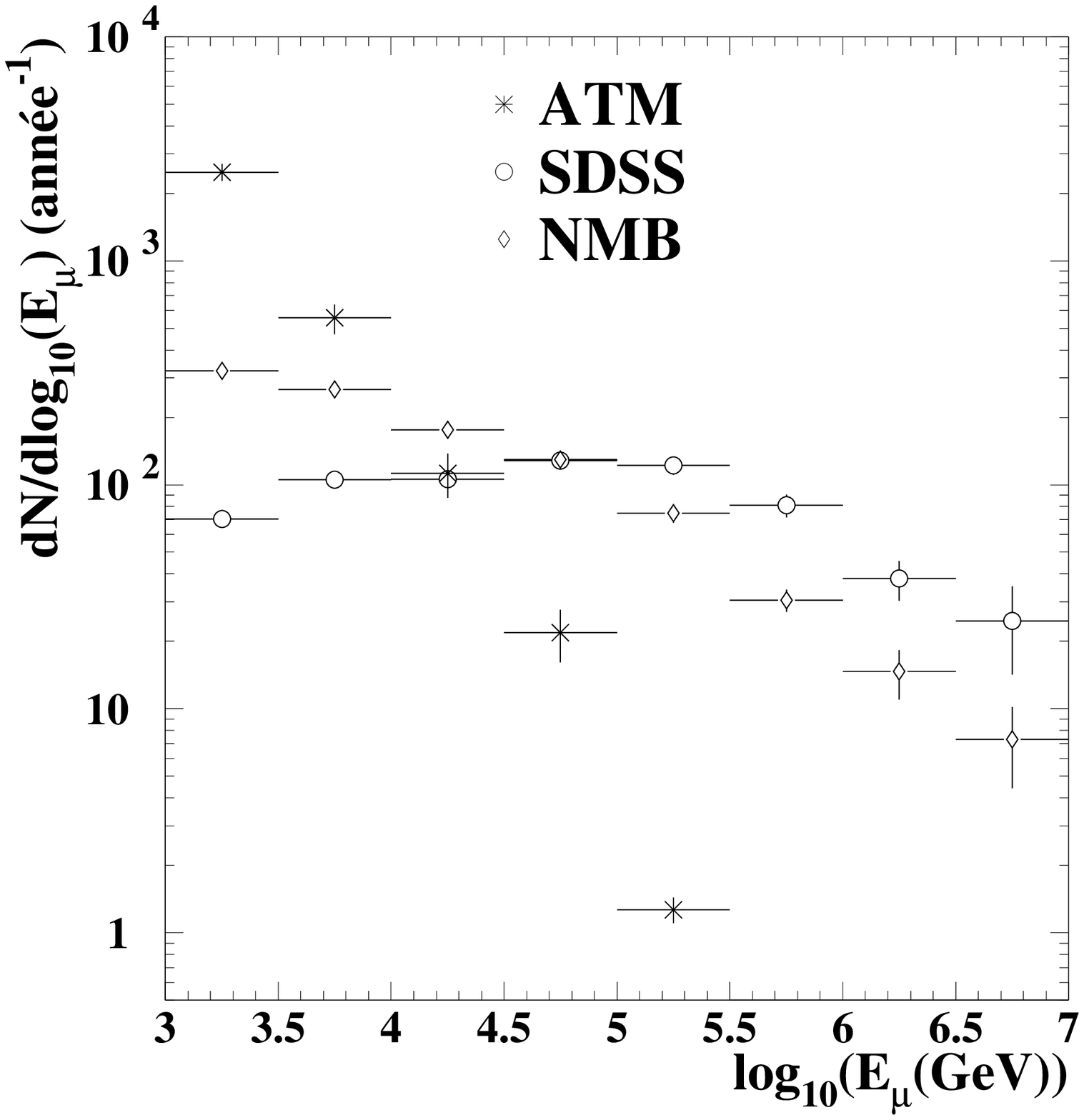, width=0.495\linewidth}
\epsfig{figure=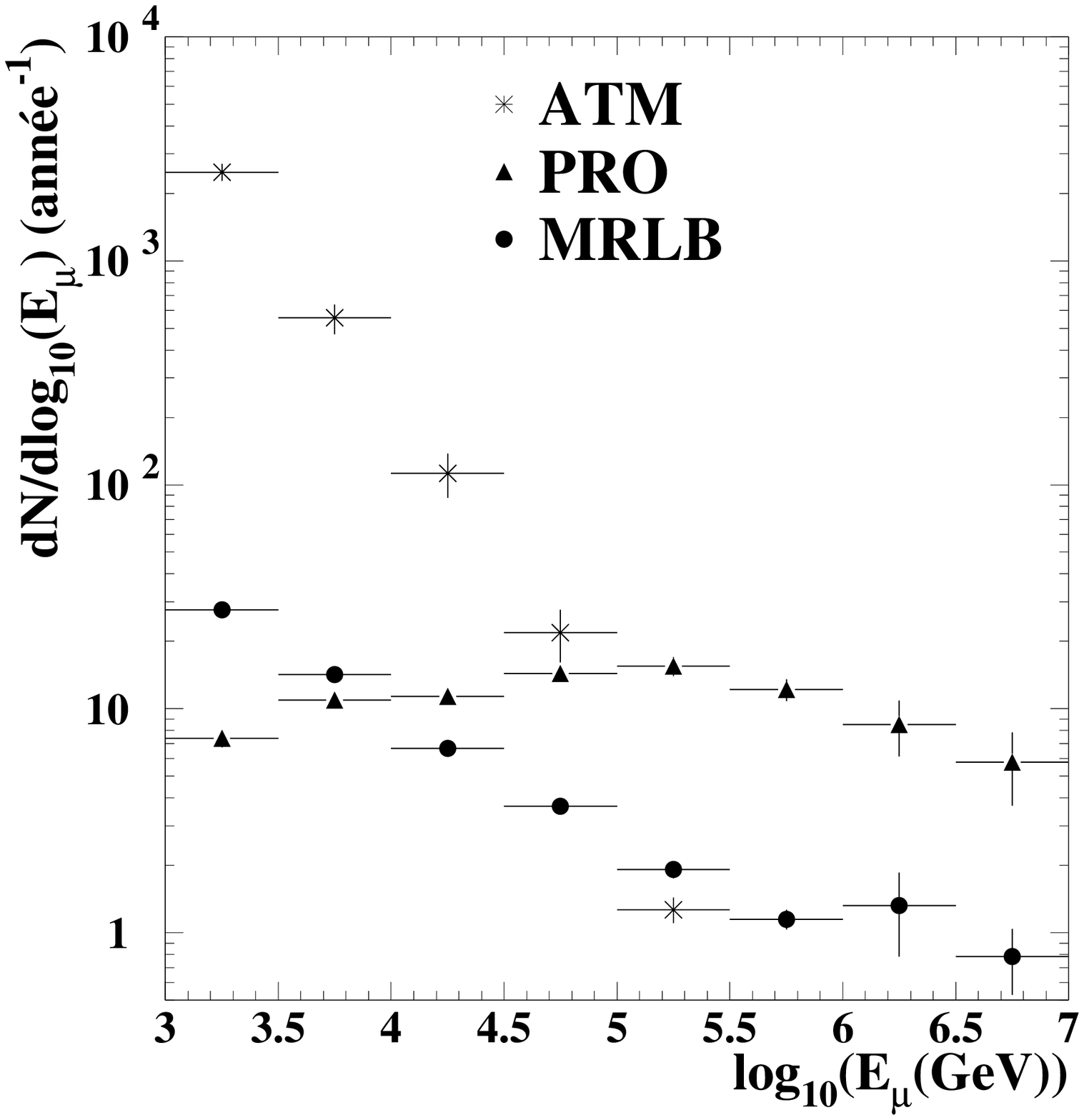, width=0.495\linewidth}
\caption{Reconstructed energy spectrum of muons issued from neutrinos. A 0.1 km$^2$ detector composed of 1000 photomultipliers has been considered. For the models NMB \cite{ref6} and SDSS \cite{ref7}, the neutrinos are produced in the vicinity of the massive black hole. For the models PRO \cite{ref8} and MRLB \cite{ref9}, they are produced in the jets. The atmospheric neutrinos ATM  \cite{ref10} which is an irreducible background is also displayed. The error bars are only statistical.}
\label{fig:all_spec_rec}
\end{figure}

   As expected, the energy resolution of such a detector is poor at high energy \hbox{($>$ TeV):}
\begin{itemize}
  \item A precision of a factor 3 can be obtained below 10 TeV,
  \item A precision of a factor 2 beyond 10 TeV.
\end{itemize}
This precision is sufficient to reconstruct the energy spectrum and to apply the energy cut necessary to suppress the atmospheric neutrinos background  (figure \ref{fig:all_spec_rec}).

\begin{figure}
\centering
\epsfig{figure=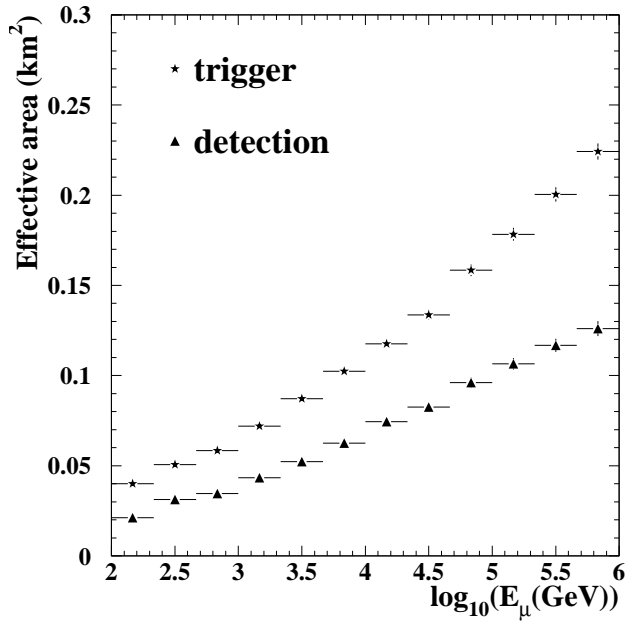, width=9cm}
\caption{Triggered and reconstructed effective surface as a function of the muon energy. The surfaces are averaged over the whole lower hemisphere and the errors bars are only statistical.}
\label{fig:area}
\end{figure}

  Moreover the detection efficiency of the detector increases with the energy (figure \ref{fig:area}). An effective surface of 0.055 km$^2$ is reached at 10 TeV and  0.1 km$^2$ beyond 100 TeV. Such a large surface leads to a measurable neutrinos fluxes as displayed on the table \ref{tab:numofevents}. It is also possible to see that the number of atmospheric neutrinos can be kept at a reasonable level. For the atmospheric muons, only an upper limit has been obtained so-far (less than 1000 events per year) due to limitations on the CPU time. However computer simulations are pursued and this limit will be improved.

\begin{table}[h]
\begin{center}
\begin{tabular}{*{3}{l}}
\hline
{\bf AGNs models} & E$_\mu^{\mathrm{rec}} > 10$~TeV &  E$_\mu^{\mathrm{rec}} > 100$~TeV\\
\hline
SDSS 	    & 251$\pm$12                      & 134$\pm$10                  \\
NMB 	    & 217$\pm$9                       &  64$\pm$4                   \\
PRO         &  34$\pm$2                       &  21$\pm$2                   \\
MRLB        & 7.8$\pm$0.4                     & 2.6$\pm$0.3                 \\
\hline
            &                                 &                             \\
\hline
{\bf Background source} & E$_\mu^{\mathrm{rec}} > 10$~TeV &   E$_\mu^{\mathrm{rec}} > 100$~TeV\\
\hline
Atmospheric neutrinos   &  68$\pm$13          & 0.8$\pm$0.1     \\
\hline
\end{tabular}
\vspace{3mm}
\caption{Number of detected events in one year with a 0.1 km$^2$ deep sea detector for different AGNs models (diffuse neutrinos source). The cut on energy is directly applied to the reconstructed energy of the muon. In the SDSS \cite{ref7} and NMB \cite{ref6} models the neutrinos are produced in vicinity of the black hole and in the PRO \cite{ref8} and MRLB \cite{ref9} models in the jets. The number of atmospheric neutrinos is also indicated. For the atmospheric muons, only an upper limit have been obtained (less than 1000 events per year) due to the present limitations on the CPU time available for the simulation. The indicated errors are statistical.}
\label{tab:numofevents}
\end{center}
\end{table}

\newpage

  \subsection{Astronomical potentialities}

  Given the detector latitude (43$^\circ$ North), about 3.5 $\pi$ steradian can be observed thanks to the Earth rotation (such a detector works all day long and observes continuously half of an hemisphere). The declinations below -47$^\circ$ are always observable, part of the day between -47$^\circ$ and +47$^\circ$, never above +47$^\circ$. Such an observable sky is particularly convenient: about 300$^\circ$ of the galactic plane (including the galactic center) is visible most of the sideral day. An other convenient feature is the fact that there will have an overlap between ANTARES and AMANDA \cite{ref15} in the South pole which observes between +90$^\circ$ and 0$^\circ$.

  With the expected performances, a 0.1 km$^2$ detector should be able to observe diffuse neutrinos sources. For point-like sources, the situation should be also favourable. Due to the very good angular resolution, the  observable sky will be covered by about 200000 pixels\footnote{This number of pixel is computed by supposing an angular resolution of 0.2 degrees. However by using the source position in the sky, a better angular resolution could be achieved for point-like sources. Therefore the sky could be covered by a larger number of pixel and the detection of point-like source easier.}. The mean number of background events in one pixel will not exceed 1.3$\cdot 10^{-2}$ per year and an excess of about 3 neutrinos inside a pixel will be sufficient to observe a point-like source.

\subsection{Study of the Neutrino Oscillation}

 The ANTARES collaboration has studied the possibility to cross-check the SuperKamiokande result \cite{ref1} by using the method described in the section {\it 2.1.4}. Results are yet preliminaries and more work is needed. Only the most important features will be given here.

  First of all a shorter vertical spacing of the Optical Modules is necessary to have a better sampling of the muon track (typically a distance of 8 meters between Optical Modules is necessary). However it appears that the optimal geometry for neutrino oscillation is not very far from the one optimal for astronomy. Concerning the angular resolution, this one is dominated by the physical process, not really by the track reconstruction due to the low muon energy (between 5 and 60 GeV). For example, by studying vertical events seen by only one string, half of the events have an angular error between the neutrino and the muon better than 1.8$^\circ$. However, if the events seen by two strings are used, we are sensitive to events having a larger energy and the angular resolution reaches 0.4$^\circ$.

  Finally the background is mainly due to the external $\nu_\mu$ which are reconstructed as contained. The contribution of the atmospheric muons and of the contained $\nu_e$ is negligible.  Even if more work is necessary, the background seems to be handled and a 0.1 km$^2$ detector as foreseen by the ANTARES collaboration should be able to explore the region of SuperKamiokande signal in a few years running.

\section{Conclusion}

  Since its beginning in Spring 1996, considerable efforts have been done by the ANTARES collaboration to design a deep underwater neutrino detector. During the R\&D phase of the project, several autonomous set-up have been built and deployed many times to measure the environmental parameters. In parallel a string prototype has been conceived to study the problems of deployment and operating of such a structure. Finally extensive computer simulations have been pursued to optimise the detector geometry and to estimate its performances. It appears that a 0.1 km$^2$ should be able to observe diffuse sources and several point-like sources.

  Now, the ANTARES collaboration aims at building a 0.1 km$^2$ detector. Its performances should be sufficient to study diffuse neutrinos sources and to observe the nearest point-like sources. The study of neutrino oscillation could be also feasible. The very first string would be deployed at the end of 2000 and the whole detector completed by the end of 2003. 
  
  Such a detector could open a new window on the Universe, in the same way as Galileo's first telescope did some 400 years ago...

\section*{Acknowledgements}

I would like to thank F. Bernard, C. C\^arloganu, F. Hubaut and S. Navas for their help and their support, and L. Moscoso for fruitful discussions.


\section*{References}

\begin{thebibliography}{99}

\bibitem{ref1} Y. Fukuda et al., {\it Phys. Rev. Lett.} {\bf 81}, 1562 (1998).

\bibitem{ref2} E. Waxman and J. Bahcall, High Energy Neutrinos from Astrophysical Sources: an Upper Bound, hep-ph/9807282 (1998).

\bibitem{ref3} K. Mannheim, R.J. Protheroe and J. Rachen, On the Cosmic Ray Bound for Models of Extragalactic neutrino Production, hep-ph/9812398 (1998).

\bibitem{ref4} E. Waxman and J. Bahcall, High Energy Astrophysical Neutrinos: the Upper Bound is Robust, hep-ph/9902383.

\bibitem{ref5} E. Waxman and J. Bahcall, {\it Phys. Rev. Lett.} {\bf 78}, 2292 (1997).

\bibitem{ref6} L. Nellen, K. Mannheim and P.L. Biermann, {\it Phys. Rev.} {\bf D47} 5270 (1993).

\bibitem{ref7} F.W. Stecker, C. Done, M.H. Salamon and P. Sommers, High Energy Neutrinos from Active Galactic Nuclei, {\it  Phys. Rev. Lett.} {\bf 66} 2697 (1991), Errata {\it Phys. Rev. Lett.} {\bf 69} 2738 (1992).

\bibitem{ref8} R.J. Protheroe, High Energy Neutrinos from Blazars, ADP-AT-96-7 and astro-ph/9607165 (1996).

\bibitem{ref9} K. Mannheim, High Energy Neutrinos from Extragalactic Jets, {\it Astropart. Phys.} {\bf 3}  295 (1995).

\bibitem{ref10} L.V. Volkova, Energy Spectra and Angular Distributions of Atmospheric Neutrinos, {\it Sov. J. Nucl. Phys.} {\bf 31} 784 (1980).

\bibitem{ref11} S. Yoshida, H. Dai, C.H. Jui and P. Sommers, Extremely High Energy Neutrinos and their Detection, astro-ph/9608186 (1996).

\bibitem{ref12} ANTARES Web page: http://antares.in2p3.fr/antares 

\bibitem{ref13} IFREMER Web page: http://www.ifremer.fr/

\bibitem{ref14} F. Hubaut, Th\`ese de Doctorat, Optimisation et caract\'erisation des performances d'un t\'elescope sous-marin \`a neutrinos pour le projet ANTARES, {\it Universit\'e de la M\'edit\'erran\'ee, Aix-Marseille II} (April 1999).

\bibitem{ref15} AMANDA Web page: http://amanda.berkeley.edu/ 

\end{thebibliography}
\end{document}